\documentclass[pre,twocolumn,
showpacs,  
amsmath]{revtex4}
\usepackage{graphicx,amssymb}

\begin{document}
\title{Prevalence of marginally unstable periodic orbits in chaotic 
billiards}
\date{\today}
\author{E. G. Altmann$^1$, T. Friedrich$^2$, A. E. Motter$^3$, H. Kantz$^1$, and A. Richter$^2$}

\affiliation{
1 Max-Planck-Institut f\"ur Physik komplexer Systeme, 01187 Dresden, Germany\\
2 Institut f\"ur Kernphysik, Technische Universit\"at Darmstadt, 64289 Darmstadt, Germany\\
3 Department of Physics and Astronomy, Northwestern University, Evanston, IL 60208, USA
}

\begin{abstract}
The dynamics of chaotic billiards
is significantly influenced by coexisting regions of regular motion. Here
we investigate the prevalence of a different fundamental structure,
which is formed by marginally unstable periodic orbits and stands apart from the regular regions.
We show that these structures both {\it exist} and {\it strongly influence} the 
dynamics of locally perturbed billiards, which include a large class of widely
studied systems. We demonstrate the impact of these structures in the quantum regime using microwave experiments in annular billiards.
\end{abstract}

\maketitle

\section{Introduction}\label{sec.I}

Chaotic billiards are fundamental paradigms in statistical physics and
nonlinear dynamics. By connecting dynamics with geometry, billiards serve
as models to address numerous questions ranging from the foundations of the ergodic
hypothesis~\cite{sinai2,stadium2} and the description of shell effects~\cite{brack} to the design of microcavity lasers \cite{hui}
and microwave resonators~\cite{ima,stoeckmann}, among other applications~\cite{stoeckmann}.

\begin{figure}[!ht]
\includegraphics[width=1\columnwidth]{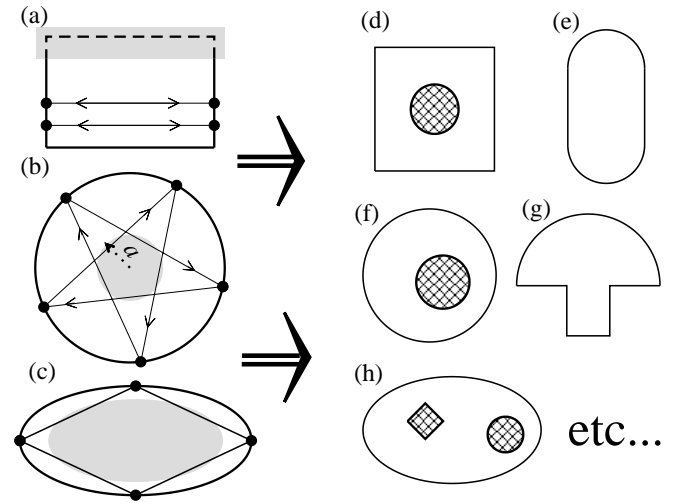}
\caption{ Adding {\em local} perturbations to integrable billiards, as
    those shown in (a)-(c), one obtains frequently studied chaotic billiards, such as
    those shown in (d)-(h). The gray regions of the (a) rectangular, (b)
    circular, and (c) elliptical billiards are defined in  such a way
    that chaotic motion is possible ($a<R$ in (b) is the radius of the
    smallest circle that circumscribes all scatterers). 
MUPOs are shown here to exist in billiards such as (d) Sinai~\cite{sinai2}, (e) 
  stadium~\cite{stadium2}, (f) annular~\cite{saito}, (g) mushroom~\cite{bunimovich}, and (h) elliptical with scatterers~\cite{note2}.}\label{fig.compa}
\end{figure}

A salient feature of billiard systems is that simple geometries, such as those in Fig.~\ref{fig.compa}, suffice to give rise to a
rich variety of dynamical behavior observed in typical Hamiltonian systems. But as previously observed for specific chaotic
billiards, simple geometries may also lead to the existence of  the so-called {\em   bouncing-ball orbits}:  one-parameter
families of periodic orbits exhibiting perpendicular motion between parallel
walls.
Theoretical and experimental work on the Sinai
[Fig.~\ref{fig.compa}(c)] and Bunimovich stadium [Fig.~\ref{fig.compa}(d)] billiards
have shown that such orbits have a major influence on transport properties,
decay of correlations, and spectral properties~\cite{stoeckmann,parallel,stadium,transport}.
This is so because, contrary to the other orbits embedded in the chaotic
component of the phase space,
bouncing-ball orbits are only marginally unstable (i.e., perturbations grow only linearly in time).
In general, marginally unstable periodic orbits (MUPOs) can be regarded as a
source of
regular behavior that masks strong chaotic properties. However, MUPOs are not structurally stable and may be destroyed by small changes in the parameters of the system. Therefore, MUPOs are considered to be non-generic and it has long been assumed that they could exist only for very special systems, like billiards with parallel walls.

Contrary to this expectation, in this paper we show that MUPOs are prevalent in a large
class of billiard systems. The starting point of our analysis is the observation that
many of the most widely studied chaotic billiards consist of {\it local} perturbations of an integrable
billiard. For concrete examples, consider the chaotic billiards shown in the right part of
Fig.~\ref{fig.compa}. All these billiards can be obtained by re-defining the
dynamics in the gray region of the integrable billiards in
Figs.~\ref{fig.compa}(a)-(c), e.g., by introducing a scatterer. It can
be shown that any orbit ({\it i}) lying inside the chaotic component and ({\it ii})
not interacting
with the introduced scatterers will be a MUPO.
Although bouncing-ball orbits evidently satisfy these conditions
in the billiards of Figs.~\ref{fig.compa}(d)-(h),
the existence of such orbits is far from clear in general.
Here we use geometric and analytical arguments to demonstrate the widespread occurrence of MUPOs.
Specifically, using circular-like billiards  as model
systems --- such as those in Figs.~\ref{fig.compa}(f)-(g) ---  we show that
{\em infinitely} many families of MUPOs exist for almost all parameter choices of
the system. We discuss the impact of these structures on the dynamics of chaotic orbits
as well as the experimental observation of MUPOs in the quantum spectrum of
microwave annular billiards.

The {\it local} perturbations described above are typical for 
billiard systems and differ fundamentally from the {\em global} perturbations
considered in smooth Hamiltonian systems. 
In the latter, the KAM theory shows that most quasi-periodic orbits of the
integrable system survive the
perturbation, while all periodic orbits with marginal stability disappear.
Quite the opposite happens in the former case: a large set of quasi-periodic orbits disappears
but there are families of periodic orbits with marginal stability that survive the perturbation
by ``avoiding'' interaction with the localized scatterers.
These orbits give rise to families of MUPOs detached from regular
regions, which were previously observed in billiards with parallel
walls~\cite{parallel}, and for specific parameters of the mushroom
billiard~\cite{nossos,nossos2}. Here we consider {\em generic} control parameters of a wide
class of systems where we characterize the MUPOs both theoretically and
experimentally.

The paper is organized as follows. In Sec.~\ref{sec.II} we perform a detailed
analysis of the existence of MUPOs in the annular billiard, a representative
example of the class of billiards we are interested in. In Sec.~\ref{sec.III} we
show the existence of an infinite number of different families of MUPOs in annular
and other circular-like billiards. Our experimental results on microwave
cavities appear in Sec.~\ref{sec.IV}. Finally, our conclusions are summarized
in Sec.~\ref{sec.V}.

\section{Annular billiard}\label{sec.II}

\begin{figure}[!ht]
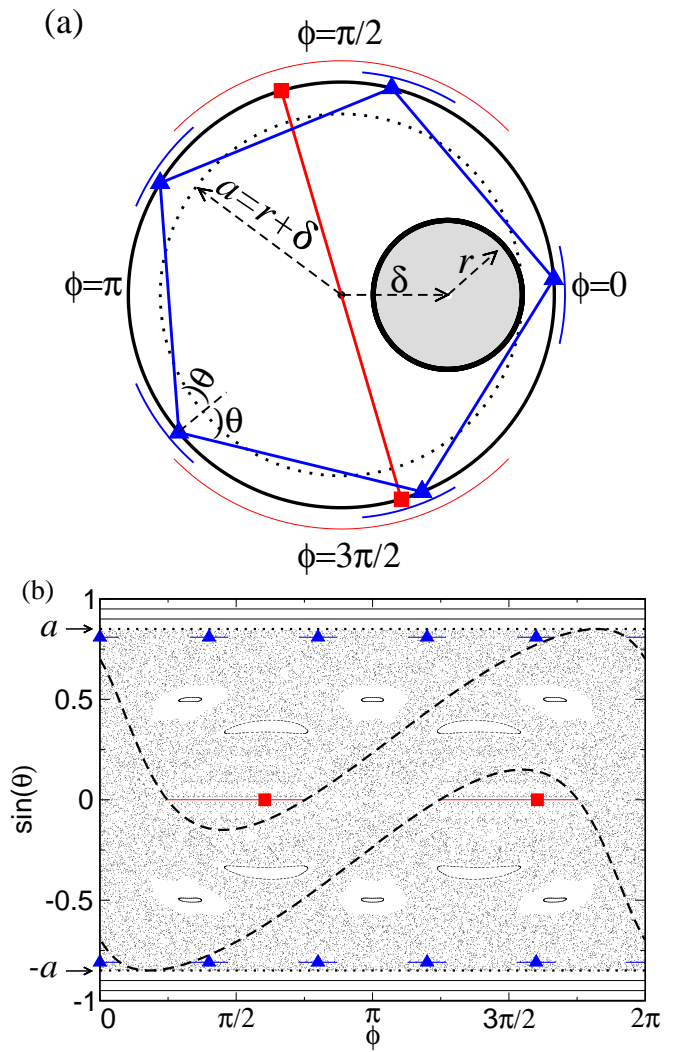

\includegraphics[width=0.9\columnwidth]{fig2a.eps}\\
\includegraphics[width=1\columnwidth]{fig2b.eps}
\caption{(Color online) Annular billiard for parameters $r=0.35$ and $\delta=0.5$:
  (a) configuration space and (b) phase space. MUPOs correspond to the periodic
  orbits that cross the circle of radius $a$ [dotted line in (a)]  but that do
  not hit the scatterer [region between the dashed lines in (b) in which
  relation~(\ref{eq.hitting}) is satisfied]. 
  The symbols {\tiny $\blacksquare$} and {\small $\blacktriangle$} indicate,
  respectively, individual orbits belonging to MUPOs~$(2,1)$ and~$(5,1)$.}\label{fig.configuration}
\end{figure}

Annular billiards are defined by two eccentric circles, as shown in
Fig.~\ref{fig.configuration}(a). For a fixed radius $R=1$  of the external circle,
the control parameters are the radius $r$ and displacement $\delta<1-r$ of the internal
circle, which serves as a scatterer.
The phase space shown in
Fig.~\ref{fig.configuration}(b) is obtained by plotting the position $\phi\in[0,2\pi]$ of the collision of
the particle with the external circumference and the sine of the angle~$\theta\in[-\pi/2,\pi/2]$ with the normal
direction right after the collision.
In this system, periodic orbits of period~$q$ and
rotation number~$\eta$ that collide only with
the external circumference define {\em star polygons} of type~$(q,\eta)$, where the
integers $q$ and $\eta$
are coprime and $\eta \leq q/2$.
A star polygon of type~$(5,2)$ is shown in Fig.~\ref{fig.compa}(b) and star polygons of
types~$(2,1)$ and~$(5,1)$ are shown in Fig.~\ref{fig.configuration}. 
Each star polygon belongs to a family of orbits of the same type, which is
parameterized by~$\phi$ and has a
fixed collision angle
$\sin(\theta_{sp})=\cos(\pi\eta/q)$.

For this system, the conditions ({\it i}) and ({\it ii}) for the existence of
MUPOs mentioned in Sec.~\ref{sec.I} translate into crossing the
circle of radius $a=r+\delta$
without colliding with the scatterer.
Under these conditions, the orbits are embedded into the chaotic sea but are only marginally unstable (both eigenvalues
of the Jacobian matrix equal $1$). The two orbits shown in  Fig.~\ref{fig.configuration} satisfy these conditions
and hence are MUPOs. We use MUPOs~$(q,\eta)$ to denote the entire
one-parameter {\em family}
of orbits corresponding
to star polygons~$(q,\eta)$ that satisfy conditions ({\it i}) and ({\it ii}). Note that these orbits are necessarily
periodic because non-periodic orbits will either collide with the scatterer or
form a regular region for $|\sin(\theta)|>a$, called
whispering gallery. A collision with the scatterer happens
whenever~\cite{saito}
\begin{equation}\label{eq.hitting}
|\sin(\theta) -\delta\sin(\theta-\phi)| > r.
\end{equation}
In Fig.~\ref{fig.configuration}b this condition is satisfied between the
dashed lines. In the following we calculate the geometrical conditions for the
existence of MUPOs and we demonstrate that typically an infinite number of
families~$q,\eta$ satisfy these conditions.

Consider MUPOs that encircle the scatterer from outside, like the
pentagon-MUPO~$(5,1)$ in Fig.~\ref{fig.configuration}. Conditions for the
existence of such {\em outer} MUPOs are obtained by
noting that every star polygon $(q,\eta)$ draws an inner regular $q$-sided
polygon, like the pentagon in Fig.~\ref{fig.compa}(b).
The radii $(d,D)$ of the inscribed and circumscribed circles of this inner
polygon are given by
$d= \cos(\pi \eta/q)$ and $D= d/\cos(\pi/q)$.
It follows that an orbit
of type~$(q,\eta)$ is an outer MUPO~$(q,\eta)$ if and only if
\begin{equation}\label{eq.ineq2}
\cos(\pi \frac{\eta}{q})< \cos( \pi \lambda) \; <\;
\frac{\cos(\pi\eta/q)}{\cos(\pi/q)}+r(1-\frac{1}{\cos(\pi/q)}),
\end{equation}
where $\cos(\pi \lambda) \equiv a=r+\delta$. A similar expression is obtained
for mushroom billiards~\cite{nossos,thesis}.

\begin{figure}[!ht]
\includegraphics[width=1\columnwidth]{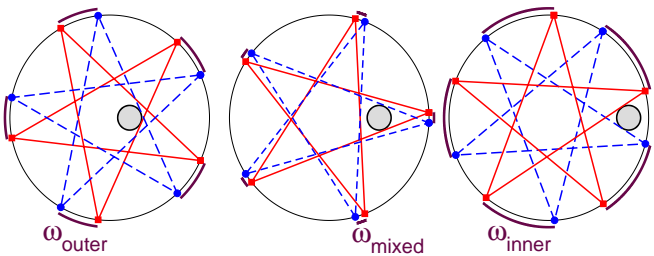}
\caption{(Color Online) Size~$w$ of the families of MUPOs~$(5,2)$ in the annular billiard with~$r=0.12$: (a)
  outer MUPOs for~$\delta=0.2$, (b) mixed inner-outer MUPO for~$\delta=0.5$,
  and (c) inner MUPOs for~$\delta=0.8$. All three kinds of MUPOs
  may coexist for a fixed~$\delta$. The size~$w$ of the families of
  MUPOs in Eq.~(\ref{eq.w}) is given by the length of the external arcs.
 MUPOs~$(5,2)$ outside of these regions do not exist since they collide with the inner scatterer.}\label{fig.new}
\end{figure}

{\em Inner} MUPOs, like the diameter-MUPO~$(2,1)$ in Fig.~\ref{fig.configuration}, exist when
$$
\delta >
\frac{r}{\cos(\pi (1-\eta)/q)}+\cos(\pi \frac{\eta}{q})+\sin(\pi \frac{\eta}{q})
\tan(\pi \frac{1-\eta}{q}). 
$$
{\em Mixed} inner-outer MUPOs may also exist for~$\eta\geq2$.  Families of inner, outer, and
mixed MUPOs~$(5,2)$ are illustrated in Fig.~\ref{fig.new}.

Unlike the regular regions around
stable periodic orbits, MUPOs have zero Lebesgue measure in the phase
space. The relevant measure is therefore the size of the families of MUPOs, given by the length~$w$
of the set of angles~$\phi$ (normalized by 2$\pi$) for which an orbit with
a given $(q,\eta)$ exists.  In Fig.~\ref{fig.configuration}, $w$ is proportional to the length of
the external arcs of  circumference in (a) and to the horizontal lines in (b). For orbits inside the whispering
gallery one would have~$w=1$. For a given family of MUPOs~$(q,\eta)$ we calculate~$w$ as 
\begin{equation}\label{eq.w}
w=w_{outer}+w_{inner}+w_{mixed} < 1,
\end{equation}
where $w_{outer}=1-q \beta^-/\pi$  and $w_{inner}= q \beta^+/\pi$
with $\cos(\beta^\pm)=[\cos(\pi \eta/q)\pm r]/\delta$. For the
MUPOs~$(q,\eta)$ investigated in Sec.~\ref{sec.IV} below, $w_{mixed}=0$. Figure~\ref{fig.new}
shows the geometrical representation of the terms in Eq.~(\ref{eq.w}).

\section{Infinite families of MUPOs}\label{sec.III}

We now determine the number and values of the {\em different} families of
MUPOs $(q,\eta)$'s that exist in a given annular billiard~$(r,\delta)$. For
inner and mixed MUPOs, only a finite number of $(q,\eta)$'s
exist~\cite{note1}, which can be obtained
by inspection.
On the other hand, we show next that an infinite number of outer MUPOs~$(q,\eta)$ typically accumulate close to the whispering gallery. 
Let~$\eta(q)$ denote the integer~$\eta$ for which
$\eta/q-\lambda$ is minimal and non-negative.
In the limit $q \rightarrow \infty \Rightarrow (\frac{\eta(q)}{q} -
\lambda)  \rightarrow 0_+$, both inequalities~(\ref{eq.ineq2}) are satisfied if
\begin{equation}\label{eq.varepsilon}
 \frac{\eta(q)}{q} - \lambda <  \frac{a \pi}{2 \sqrt{1-r^2}} \frac{1}{q^2}.
\end{equation}
Essentially the same expression is obtained for mushroom
billiards~\cite{thesis} and the same scaling on~$q$ is expected in the case of
other circular-like billiards~\cite{note2}.

Optimal rational approximants
of~$\lambda=\arccos(a)/\pi$ for a fixed~$q$ are obtained by truncating the continued
fraction representation $\lambda =
\frac{1}{\alpha_1+\frac{1}{\alpha_2+...}}=[\alpha_1,\alpha_2,...]$, leading to
the convergent~$\eta'/q'$.
The irrational numbers $\lambda^*$ for which there exists one
integer~$\alpha_{\mathrm{max}}$ such that
$\alpha_i<\alpha_{\mathrm{max}}$, for all $i$,
are called {\em numbers of constant type}.
Numbers of constant type  are
difficult to approximate by rational numbers and
there exist constants~$C_1,C_2$ such that~\cite{khinchin}
\begin{equation}\label{eq.c1c2}
\frac{C_1}{q^2} <  \left| \frac{\eta'}{q'} - \lambda^* \right| < \frac{C_2}{q^2},
\end{equation}
for all convergents~$\eta'/q'$. Comparing the 
inequalities~(\ref{eq.varepsilon}) and~(\ref{eq.c1c2}) we note the
same~$q^{-2}$ dependence.
Since the
convergents are the {\em best} approximants, the lower bound
in~(\ref{eq.c1c2}) is valid for all rational numbers. Therefore,
provided that $\lambda$ is a number of constant type, there are regions of the
control parameters [$a\pi/(2\sqrt{1-r^2})<C_1$ for
annular billiards] for which there exist only a finite number of
families of MUPOs.
The numbers of constant type are uncountable and dense in
the set of real numbers. They have zero Lebesgue measure, however, meaning
that with full probability~$\lambda$ belongs to the complementary
set of irrational numbers for which $C_2\rightarrow0$ in Eq.~(\ref{eq.c1c2}). Therefore, an
infinite number of MUPOs exist for almost all~$\lambda$ and hence for almost
parameters~$(r,\delta)$. 

The demonstration above can be used in circular-like
billiards with arbitrary inner scatterers~\cite{note2} to verify 
whether the convergents $\eta'/q'$ of~$\lambda=\arccos(a)/\pi$ are MUPOs~$(q',\eta')$
[e.g., satisfy condition~(\ref{eq.ineq2}) in the case of annular billiards or
 Eq.~(6) of Ref.~\cite{nosso.mushroom} in the case of mushrooms].
Typically, an infinite number of different families~$(q,\eta)$ can be found among the
convergents.
For the annular
billiard illustrated in Fig.~\ref{fig.configuration}, for instance, all odd convergents tested are MUPOs:
$(5,1),(11,2),(436,77),(1342,237),...$, while the MUPO~$(4,1)$ is not a
convergent.

\section{Experimental results}\label{sec.IV}

Having shown that MUPOs are abundant, we now study the impact of these structures
in quantum experiments. We use the equivalence between Schr\"{o}dinger's and Helmholtz's equations for flat
microwave cavities \cite{ima,stoeckmann} to investigate the effect of MUPOs
in quantum annular billiards. We show that MUPOs are detectable and play a
prominent role among the periodic orbits.

A microwave cavity with radius $12.5\,$cm, height of $5\,$mm, and $4$
coupling antennas was used in the experiments. The inner scatterer
had a radius of $1.5\,$cm, leading to~$r=0.12$. The resonance
spectra have been obtained using a vectorial network analyzer,
measuring the complex amplitude ratio of the input and output
microwave signal of the cavity. For each value of $\delta=0,\,0.08,
\cdots, 0.88 $ we measured $10$ spectra up to $10\,$GHz with a resolution of
$100\,$kHz. Different antennas and antenna combinations were used to find as
many resonances as possible.
Close lying levels (e.g., split doublets) were detected as
one resonance only due to their finite width. However, since the
position of those doublets can be approximately calculated, a
second (\emph{not detected}) eigenvalue could be attributed to the
corresponding frequencies. We justify this procedure by using the
high precision data obtained with superconducting cavities in the
experiments described in Ref.~\cite{exp}: there the doublets could
be resolved, and we found that the length spectrum is stable under
small random shifts of one doublet partner, which allows us to
assume doublets to be degenerate. Finally, by comparing the number
of detected levels $N(f)$ below the frequency $f$ to the expected
number $N_\mathrm{Weyl}$ given by Weyl's formula \cite{brack}, we
checked that almost all eigenvalues in the considered part of the
spectrum have been found ($N \approx 150$ for each
$\delta$).

\begin{figure}
\includegraphics[width=1\columnwidth]{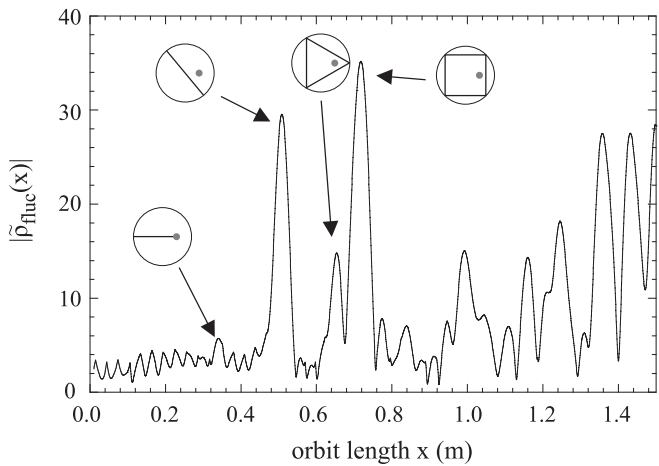}
\caption{\label{fig_lsp} Experimental length spectrum of the annular billiard
for $\delta=0.48$. Peaks associated to four periodic orbits are indicated: the
shortest one is unstable, the diameter and triangle are MUPOs, and the square
is inside the regular region.} 
\end{figure}

Performing a Fourier transform ($\mathcal{FT}$) of the level density
$\rho(f)=\frac{dN(f)}{df}$ we have computed the length spectrum
\begin{equation}
  |\tilde{\rho}_{fluc}(x)|=|\mathcal{FT}\{\rho(k)-\rho_{\mathrm{Weyl}}(k)\}|\,,
\label{eq_lsp}
\end{equation}
where $k=2\pi f/c$. The classical periodic orbits manifest themselves as peaks located at
the corresponding orbit length. The length of periodic orbit~$(q,\eta)$ is
given by~$x_{(q,\eta)}=2Rq\sin(\pi \eta/q)$. Particularly, for
all~$\delta$'s we consider the peak heights~$y$ (strengths) of the
diameter ($x_{(2,1)}=0.5\,$m), triangular ($x_{(3,1)}=0.63\,$m),
and square ($x_{(4,1)}=0.71\,$m) orbits. The length spectrum
for $\delta=0.48$ is shown in Fig.~\ref{fig_lsp}, where we indicate
additionally the peak at $x=0.34\,$m related to an {\em unstable}
periodic orbit. Notice that this peak is much smaller than the
peaks associated with the MUPOs.

\begin{figure}
\includegraphics[width=1\columnwidth]{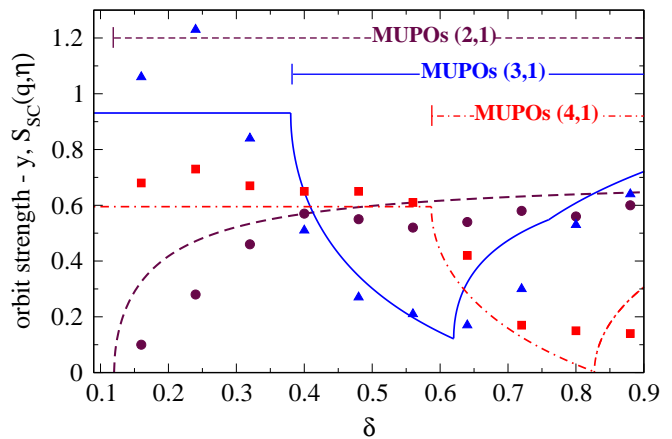}
\caption{\label{fig_comparison} (Color online) Semiclassical strengths
  $S_{sc}$  (lines) compared to the
  experimental strengths $y$ (symbols). $S_{sc}$ is given by
  Eq.~(\ref{eq.ssc}) and expected from periodic orbit theory, while $y$ is
  extracted from the length spectra, as shown in Fig.~\ref{fig_lsp}. Three orbits are considered: diameter
  (dotted line and circles), triangle (solid line and triangles), and 
square (dot-dashed line and squares). The horizontal lines (top) indicate the
  values of~$\delta$ for which the corresponding MUPOs exist.}
\end{figure}

Using periodic orbit theory, the strength of an orbit in a quantum
mechanical length spectrum is given by the amplitudes of the
oscillatory terms in a semiclassical periodic orbit summation.
We use the trace formula for integrable systems to obtain the orbit dependent amplitudes
$\mathcal{A}=\nu\,\frac{\sin^{3/2}{(\pi\eta/q)}}{\sqrt{q}}\,,$
where $\nu=1$ for the diameter orbit, and $\nu=2$ for all other
orbits~\cite{brack}. The expected strength in the case
of the MUPOs is 
\begin{equation}\label{eq.ssc}
S_{sc}(q,\eta)=w\mathcal{A},
\end{equation}
where $w$ is the measure of the
entire family, given in Eq.~(\ref{eq.w}) and illustrated in Fig.~\ref{fig.new}. 
In Fig.~\ref{fig_comparison} we compare $S_{sc}$ (lines)
with the experimental strengths~$y$ (symbols, rescaled by a common factor) for
different values of $\delta$.  
The dependence of $S_{sc}$ on~$\delta$ is due to the factor~$w$. Overall,
the orbits strengths~$y$  approximately follow
the semiclassical behavior~$S_{sc}$ for $\delta>0.3$.
The deviations can be understood qualitatively as the experiment diverges
from the
semiclassical limit: (1) the finite
wavelengths imply a spatial uncertainty of
the order of the typical width of the peaks in the length
spectrum; (2)   the Fourier transform of
a finite spectral range generates fluctuations in the
length spectra (of the
order of $10\,$\% of the diameter peak height, as seen for $x<0.2\,$m in 
Fig.~\ref{fig_lsp}).
Nevertheless we find that quantum behavior resembles the
classical behavior in the sense that the data support the use of the weighting factors $w$ in the semiclassical
strengths in Eq.~(\ref{eq.ssc}).

\section{Conclusions}\label{sec.V}

We have demonstrated that MUPOs are prevalent and that
they must be accounted for in billiard experiments, which is a new
paradigm that advances previous conclusions drawn for specific
systems \cite{stoeckmann,nossos,stadium}. In particular, MUPOs
have not been previously observed in annular billiards, despite
many theoretical~\cite{saito,hentschel,theoretic} and
experimental~\cite{exp} studies, including detailed catalogs of
periodic orbits~\cite{gouesbet}. We have shown that annular and general
circular-like 
billiards typically have an infinite number of different families
of MUPOs in the chaotic component close to the border of the
whispering gallery. This should be contrasted with the case of billiards with
parallel walls such as Stadium and Sinai billiards, where only a finite number
of families of MUPOs exists. 

The above mentioned results can be immediately extended to
other chaotic billiards defined by local perturbations of
integrable systems and are expected to find applications in both
classical and quantum studies. Classically, general arguments on marginal
instability  can be used to show that the resulting stickiness of
chaotic trajectories to MUPOs generates a universal power law
$p(t)\sim t^{-2}$  for
the survival probability of nearby particles~\cite{nossos2}. This scaling
is expected to hold for long times, while fluctuations
(nonperiodic echoes) occur for short times~\cite{nossos,thomas}. 
Studies in the quantum regime have shown that orbits with marginal stability
are robust to small perturbations~\cite{robust} and give rise to different
transport phenomena~\cite{transport}. Recent
theoretical studies and microwave experiments on chaos assisted
tunneling in the annular billiard have demonstrated a pronounced effect on the
tunneling of the so called ``beach region'' between the whispering gallery and
the chaotic region~\cite{theoretic,exp}. 
Different mechanisms of dynamical tunneling are currently under
investigation~\cite{schlagheck}, and special attention is being devoted to
mushroom billiards~\cite{barnett,baecker}. Our results have fully characterized the
dynamics in the ``beach region'' of annular and mushroom billiards in terms
of marginal unstable orbits. The
formalization of their contribution to dynamical tunneling and a
comparison with the existing numerical and experimental results are
interesting open questions.  

\acknowledgements
We thank M. Brack, B. Dietz, and T. H. Seligman for stimulating discussions  and B. Lindner for the careful reading of the manuscript.
This work was supported by DFG  (SFB 634), Studienstiftung des Deutschen Volkes, and CAPES (Brazil).


\begin{thebibliography}{99}
\bibitem{note2} Our proof 
  of the existence of infinitely many MUPOs holds when the scatterers (1)
  induce chaotic dynamics and (2) have a
  finite number of points at the critical caustic [circle of radius~$a$ in
  Fig.~\ref{fig.compa} (b) in the case of circular-like billiards].
\bibitem{note1} The width in~$\phi$ where such star polygons may exist is
  fixed  while the distance between two 
adjacent vertices is $2\pi/q$, i.e., goes to zero with increasing~$q$ [see Fig.~\ref{fig.configuration} (b)].
\end{thebibliography}
\end{document}